# Pragmatic SAE procedure in the Schrodinger equation for the inverse-square-like potentials

**Teimuraz Nadareishvili[1,] and Anzor Khelashvili[2]**

[1] Iv. Javakhishvili Tbilisi State University Chavchavadze Ave. 3, 0162, Tbilisi, Georgia and Inst. of High Energy Physics, Iv. Javakhishvili Tbilisi State University, University Str. 9, 0109, Tbilisi, Georgia

[2] Inst. of High Energy Physics, Iv. Javakhishvili Tbilisi State University, University Str. 9, 0109, Tbilisi, Georgia and St.Andrea the First-called Georgian University of Patriarchy of Georgia, Chavchavadze Ave.53a, 0162, Tbilisi, Georgia.

*E-mail: teimuraz.nadareishvili@tsu.ge and anzor.khelashvili@tsu.ge*
Corresponding author. Phone: 011+995-98-54-47 ; E-mail :teimuraz.nadareishvili@tsu.ge

**Abstract.** The Self-Adjoint Extension in the Schrodinger equation for potentials behaved as an attractive inverse square at the origin is critically reviewed. Original results are also presented. It is shown that the additional non-regular solutions must be retained for definite interval of parameters, which requires a necessity of performing a Self-Adjoint Extension (SAE) procedure of radial Hamiltonian. The "Pragmatic approach" is used and some of its consequences are considered for wide class of transitive potentials. Our consideration is based on the established earlier by us a boundary condition for the radial wave function and the corresponding consequences are derived. Various relevant applications are presented as well. They are: inverse square potential in the Schrodinger equation is solved when the additional non-regular solution is retained. Valence electron model and the Klein-Gordon equation with the Coulomb potential is considered and the "hydrino"-like levels are discussed.

**Keywords:** Self-Adjoint Extension, singular behaviour, additional solutions, transitive potentials.

**PACS numbers:** 03.65.-w, 03.65.Ca, 03.65.Ge, 03.65.Ta

## 1. Introduction

Following to various physical requirements we have shown [1,2] that the full radial function $R(r)$ must behave at the origin as $R(r) \underset{r \to 0}{\approx} r^{-1+s}$, where $s > 0$ is an arbitrary small positive number. Behavior of this kind is a more restriction than that which follows from the finiteness of the norm. Moreover because of singularity of Laplace operator at the origin after substitution $R(r) = \frac{u(r)}{r}$ the standard equation for reduced wave function $u(r)$, consisting only a second derivative, does not follow. It appears additional term with delta function [1,2] for avoiding of which radial function is strictly restricted by the behavior $u(r) \underset{r \to 0}{\approx} r^{1+\varepsilon}$, where $\varepsilon$ is zero or positive integer according to the theory of distributions. Such behaviour takes place only for regular potentials (see, definitions below). As regards of singular potentials their consideration on the level of a reduced wave function $u(r)$ is hardly problematic and therefore we have to restrict ourself by the equation for full radial function $R(r)$. It is worthwhile that our approach has been applied in paper [3] in study of magnetic resonances between fermions and antifermions at small distances.

It is well-known that the standard reduced radial Hamiltonian

$$H_r = -\frac{d^2}{dr^2} + \frac{l(l+1)}{r^2} + 2mV(r) \qquad (1.1)$$

is usually used in case of attractive singular potentials for studying of self-adjoint extention (SAE)[4]. But it follows from our results that in such consideration without accounting above-mentioned behavior at the origin the connection with original full 3-dimensional Schrodinger equation is lost and hence the obtained results may only have mathematical interests. On the other hand the singular potentials can be considered on the basis of equation for full radial wave function $R(r)$. Among such potentials $r^{-2}$ like behaving ones attract the most attention. Number of physically significant quantum-mechanical problems manifest in such a behavior. Hamiltonians with inverse square like potentials appear in many systems and they have sufficiently rich physical and mathematical structures. Examples of such systems are: Valence electron model for hydrogen like atoms in quantum mechanics [5], the theory of black holes [6], conformal quantum mechanics [7], Aharonov-Bohm effect [8], Dirac monopoles [9], Calogero model [10],domains and spectra of the SAE of the Hamiltonian of the singular oscillator potential [11], SAE Hamiltonian in a model of supersymmetric Quantum Mechanics with SUSY breaking [12], domains of the SAE and its (non-) scale invariance, as well as the departure of the Zeta function from the case of non-singular potentials [13] and etc. Mathematical aspects of SAE in differential equations are considered also in [14].

Below we'll study SAE problems in the Schrodinger equation directly on the basis of total radial function $R(r)$. Our consideration closely follows to our earlier paper [15], in which the same boundary condition is used as here.

At small distances $r^{-2}$ like potentials have singular solutions together with regular ones. As a rule such solutions are ignored from the consideration, but by our opinion this action is not always reasonable and legitimate.

Detailed consideration of above-mentioned problems puts in doubt the motivations for neglecting of so-called additional (singular) solutions, which are based on mathematical sets of quantum mechanics without invoking of specific physical ideas.

The aim of this article is to elucidate some vague points, reviewing main papers in this direction. Original results are also presented.

In our paper we follow strictly to the vanishing boundary condition remarked above.

This paper is organized as follows: First, we bring the common reasonings under which these additional solutions are neglected usually. We show that none of them is convinced completely and this problem needs more profound investigation. In Section II we raise the problem, In Section III we show that under some circumstances it is necessary to preserve additional solutions. In Section IV SAE is introduced. In the foregoing Sections some consequences of retaining this additional (irregular) solution is discussed and various models are considered, where the problem of SAE takes place.

## 2. Statement of Problem

In this Section we briefly discuss the main properties of radial function and radial Hamiltonian in the light of above Introduction. We begin here by remembering of some definitions:

Full 3-dimensional wave function is presented as

$$\psi(\vec{r}) = R(r) Y_l^m(\theta,\varphi) \quad r > 0; \quad 0 \le \theta \le \pi; \quad 0 \le \varphi \le 2\pi \tag{2.1}$$

and satisfies 3-dimensional Schrodinger equation

$$\Delta \psi(\vec{r}) + 2m[E - V(r)]\psi(\vec{r}) = 0 \tag{2.2}$$

whereas the equation for radial function $R(r)$ takes form

$$\frac{d^2 R}{dr^2} + \frac{2}{r}\frac{dR}{dr} + 2m[E - V(r)]R - \frac{l(l+1)}{r^2}R = 0 \tag{2.3}$$

If we follow to traditional way of exlusion the first derivative term by substitution

$$R(r) = \frac{u(r)}{r}, \tag{2.4}$$

instead of well-known reduced equation, explored in the literature ([16,17] and any textbook)

, $$\frac{d^2 u(r)}{dr^2} - \frac{l(l+1)}{r^2}u(r) + 2m[E - V(r)]u(r) = 0 \tag{2.5}$$

it follows the equation with extra delta function term [1,2]

$$\frac{d^2u(r)}{dr^2} - 4\pi\delta^{(3)}(r) - \frac{l(l+1)}{r^2}u(r) + 2m[E - V(r)]u(r) = 0 \tag{2.6}$$

This unexpected fact changes many familiar things drastically. Detailed analysis shows [1,2] that for consistency of solutions in terms of reduced wave function with that of full Schrodinger equation, which is important physical requirement [17], it is necessary to impose reduced wave function by the condition $u(0) = 0$. Morever the character of tending to zero at the origin must be established carefully. We have proven that the reduced radial equation (2.5) is equivalent to the 3-dimensional Schrodinger equation (2.2) only when the function $R(r)$ has less singularity at the origin, than $1/r$ or

$$\lim_{r \to 0} rR = 0 \tag{2.7}$$

As a result it follows that the standard radial Hamiltonian (1.1) without mentioned boundary condition carries only mathematical interest. Therefore instead of it we have to consider a Hamiltonian, corresponding to full radial function $R(r)$, namely

$$H_R = -\frac{d^2}{dr^2} - \frac{2}{r}\frac{d}{dr} + \frac{l(l+1)}{r^2} + 2mV(r) \tag{2.8}$$

in the framework of mentioned boundary condition.

At the end of this Section we want remember Pauli's comment [18] in connection to this boundary condition. Pauli mentioned that $\lim_{r \to 0}(rR) = A \neq 0$ are inadmissible, while $\int_0^\infty R^*R dr$ exists; i.e. only normalizability is not sufficient. Note that ignorance of this fact are continued in the recently appeared papers as well [see, e.g. [19]), where only the square integrability is considered. Our condition folows also from the requirement of hermitisity of radial momentum operator $p_r = -i\left\{\frac{d}{dr} + \frac{1}{r}\right\}$ [20].

### 3. Problem of singular solution

Usually regular potentials are considered in the Schrodinger equation, which obey the following restriction at the origin

$$\lim_{r \to 0} r^2 V(r) = 0 \tag{3.1}$$

In this case the radial wave function behaves as [16,20]

$$\lim_{r \to 0} R = C_1 r^l + C_2 r^{-(l+1)} \tag{3.2}$$

where $l$ is orbital momentum. The second term in this expression is singular; it does not satisfy boundary condition (2.7) and should be neglected, even for $l=0$.[1]

It is also known, that for singular potentials, that behave like

$$\lim_{r \to 0} r^2 V \to \pm\infty \tag{3.3}$$

"falling to the center" takes place [22-23].

We study potentials with intermediate behavior, called "transitive potentials" or "regular-singular potentials" with

$$\lim_{r \to 0} r^2 V \to \pm V_0 \quad (V_0 = const > 0) \tag{3.4}$$

Two signs in the (3.4) correspond to repulsive (+) and attractive (-) cases, respectively.

For such potentials, the following statement can be proved:

**Theorem**. The Schrodinger equation except the standard (non-singular) solutions has also additional solutions for attractive potential, like (3.4), when the following condition is satisfied

$$l(l+1) < 2mV_0 \tag{3.5}$$

. The proof of this theorem is straightforward.

Indeed, let us consider the equation (2.3). For the attractive potential (3.4) at small distances this equation reduces to

$$R'' + \frac{2}{r}R' - \frac{P^2 - 1/4}{r^2}R = 0 \tag{3.6}$$

where

$$P = \sqrt{(l+1/2)^2 - 2mV_0} > 0 \tag{3.7}$$

Therefore, Eq. (3.6) has following solution

$$\lim_{r \to 0} R = a_{st} r^{-1/2+P} + a_{add} r^{-1/2-P} = R_{st} + R_{add} \tag{3.8}$$

So we have two regions for this parameter $P$. In the interval

$$0 < P < 1/2 \tag{3.9}$$

the second term $a_{add} r^{-1/2-P} = R_{add}$ must be also retained, because the boundary condition is fulfilled for it. The potential like (3.4) was firstly considered by K.Case [22], but he ignored the second term in solution. As regards of a region $P \geq \frac{1}{2}$, only the first term $a_{st} r^{-1/2+P} = R_{st}$ must be retained.

---

[1] **That the R-function, as the solution of the Laplace equation, does not contain negative integer powers of r was mentioned long time ago and appeared already in quantum mechanical textbooks (see, e.g. [21])**

From eqs. (3.7) and (3.9) follows the condition (3.5) of existency of additional states. If we demand the reality of $P$ (otherwise ''falling'' to center takes place [22,23]) the parameter $V_0$ would be restricted by condition

$$2mV_0 < l(l+1) + 1/4 \qquad (3.10)$$

The last two inequalities restrict $2mV_0$ in the following interval

$$l(l+1) < 2mV_0 < l(l+1) + 1/4 \qquad (3.11)$$

Intervals from the left and from the right sides have no crossing and therefore, if additional solution exists for fixed $V_0$ and for some $l$, then it is absent for another $l$.

Thus we see from (3.5) that in the $l = 0$ state except the standard solutions there are additional solutions as well for arbitrary small $V_0$, while for $l \neq 0$ the "strong" field is necessary in order to fulfill (3.5).

It should be mentioned, that additional solutions survive such traditional requirement as the normalizability of wave function [23] and the finiteness of the integral from probability density [16]. The stronger restriction on the wave function is also considered in the textbooks [24,25]. Namely, the matrix elements of kinetic energy operator are required to be finite. To this end, the average value of kinetic energy operator $T = \dfrac{<\vec{p}^2>}{2m}$ is evaluated by this additional function in $l = 0$ state for a Coulomb potential in the Klein-Gordon equation (This problem after corresponding modifications reduces to the Schrödinger equation with potential (3.4))

$$<\vec{p}^2> = \int_0^\infty \left(\frac{dR}{dr}\right)^2 r^2 dr \qquad (3.12)$$

If we calculate this expression by using $R_{add} \underset{r \to 0}{\sim} r^{-1/2-F}$, then it indeed diverges. However, in our opinion this requirement is overestimated. The finiteness of the total energy could be sufficient, and indeed, this is the case.

We can demonstrate this by using generalized virial theorem [26] just for singular potential; It differs from the usual virial theorem and can be written as

$$E = \left\langle V + \frac{1}{2}rV' \right\rangle + \frac{P^2}{m} a_{st} a_{add} \qquad (3.13)$$

where $a_{st}$ and $a_{add}$ are given by (3.8). It is evident that for "pure" standard ($a_{add} = 0$) and "pure" additional ($a_{st} = 0$) solutions the usual virial theorem follows from (3.13)

$$E = \left\langle V + \frac{1}{2}rV' \right\rangle \qquad (3.14)$$

We see that for our potential (3.4) the total energy is finite. It is clear from (3.13) that singular parts are cancelled. It is also evident, that the finiteness of total energy follows from explicit calculations as well, without using virial theorem. It will be shown below.

Thus, the total energy is finite in case under consideration and the requirement of finiteness of kinetic energy separately is very strong and unjustified.

There is an interesting remark in the book of R.Newton [27] for (3.4) like potentials. In (3.8) both terms are singular in the range (3.9). R.Newton pointed out that: "If $P < 1/2$, then the second term is non-regular in the sense that it dominates under the first one. At the same time this non-regular solution is square integrable as well and satisfies to the three – dimensional Schrödinger equation". We think that this argument does not forbid the additional solution.

To summarize all above-mentioned restrictions and comments as well as other artificial ones, we conclude that there is no satisfactory argument in the framework of quantum mechanics, which avoids this additional solution self-consistently.

Therefore, one has to retain this additional solution and study its consequences.

## 4. SAE procedure for Radial Hamiltonian in pragmatic approach

Let us remember some principal points of self-adjoint extention (SAE) procedure.

If for any functions $u$ and $v$, given operator $\hat{A}$ satisfies to the condition

$$\langle v | \hat{A} u \rangle = \langle \hat{A} v | u \rangle \tag{4.1}$$

then this operator is called hermitian (or symmetric). For self-adjointness it is required in addition that the domains of functions of operators $\hat{A}$ and $\hat{A}^+$ would be equal. As a rule, the domain of the $\hat{A}^+$ is wider and it becomes necessary to make a self-adjoint extension of the operator $\hat{A}$.

There exists a well known powerful mathematical apparatus for this purpose [28,29].

It may happen that the operator is hermitian, but its self-adjoint extension is impossible, i.e. hermiticity is the *necessary, but not sufficient condition* for self-adjointness. Good example is the operator of the radial momentum $p_r$ which is hermitian on functions that satisfy to the condition (2.7), but its extension to self-adjoint one is impossible (see, L.D.Faddeev's remark in the A.Messiah's book – Russian translation, footnote in p.336 [30]).

Our subject of interest is the radial Hamiltonian (2.8).

It is easy to see that for any two eigenfunctions $u_1$ and $u_2$ corresponding to the levels $E_1$ and $E_2$ of the radial Hamiltonian $\hat{H}_R$, the condition (4.1) takes the following form

$$\int_0^\infty R_1 \hat{H}_r R_2 r^2 dr - \int_0^\infty R_2 \hat{H}_r R_1 r^2 dr = \frac{1}{2}\lim_{r\to 0}\left[u_2(r)\frac{d[u_1(r)]}{dr} - u_1(r)\frac{d[u_2(r)]}{dr}\right] \quad (4.2)$$

Where, for convenience, we have temporarily introduced the notation

$$u_i(r) = rR_i(r); \quad i = 1,2 \quad (4.3)$$

Morever we mean only bound state solutions tending to zero at infinity and wave functions are real.

Let us now consider orthogonality relation: Write the Schrodinger equation (2.3) for arbitrary two levels $E_1$ and $E_2$:

$$\frac{d^2 R_1}{dr^2} + \frac{2}{r}\frac{dR_1}{dr} + 2m[E_1 - V(r)]R_1 - \frac{l(l+1)}{r^2}R_1 = 0 \quad (4.4)$$

$$\frac{d^2 R_2}{dr^2} + \frac{2}{r}\frac{dR_2}{dr} + 2m[E_2 - V(r)]R_2 - \frac{l(l+1)}{r^2}R_2 = 0 \quad (4.5)$$

Multiplying first equation on $R_2$, while the second – on $R_1$ and substract. By usung the known relation

$$\frac{d^2 R}{dr^2} + \frac{2}{r}\frac{dR}{dr} = \frac{1}{r^2}\frac{d}{dr}\left(r^2 \frac{d}{dr}R\right) \quad (4.6)$$

we derive

$$R_2 \frac{1}{r^2}\frac{d}{dr}\left(r^2 \frac{d}{dr}R_1\right) - R_1 \frac{1}{r^2}\frac{d}{dr}\left(r^2 \frac{d}{dr}R_2\right) + 2m(E_1 - E_2)R_1 R_2 = 0 \quad (4.7)$$

Let us now integrate this equation in spherical volume $r^2 dr$

$$\int_0^\infty \left[R_2 \frac{1}{r^2}\frac{d}{dr}\left(r^2 \frac{d}{dr}R_1\right) - R_1 \frac{1}{r^2}\frac{d}{dr}\left(r^2 \frac{d}{dr}R_2\right) + 2m(E_1 - E_2)R_1 R_2\right] r^2 dr = 0 \quad (4.8)$$

Consider the first term in more detail, integrating by parts. we obtain

$$\int_0^\infty R_2 \frac{d}{dr}\left(r^2 \frac{d}{dr}R_1\right) dr = -R_2 r^2 \frac{d}{dr}R_1 \bigg|_0^\infty - \int_0^\infty \frac{dR_2}{dr} r^2 \frac{dR_1}{dr} dr \quad (4.9)$$

Here we suppose that bound state wave functions decrease at infinity and retain only lower boundary. Analogously, for the second term in (4.8) we'll have

$$\int_0^\infty R_1 \frac{d}{dr}\left(r^2 \frac{d}{dr}R_2\right) dr = -R_1 r^2 \frac{d}{dr}R_2 \bigg|_0^\infty - \int_0^\infty \frac{dR_1}{dr} r^2 \frac{dR_2}{dr} dr \quad (4.10)$$

Taking into account the last two equations into Eq. (4.8), we derive

$$R_1 r^2 \frac{d}{dr}R_2 \bigg|_0 - R_2 r^2 \frac{d}{dr}R_1 \bigg|_0 + 2m(E_1 - E_2)\int_0^\infty R_1 R_2 r^2 dr = 0 \quad (4.11)$$

Now

$$R_1 r^2 \frac{d}{dr} R_2 \bigg|_0 = rR_1 r \frac{d}{dr} R_2 \bigg|_0 = rR_1 \left[ \frac{d}{dr}(rR_2) - R_2 \right]\bigg|_0 = rR_1 \frac{d}{dr}(rR_2) - rR_1 R_2 \bigg|_0 \qquad (4.12)$$

and

$$R_2 r^2 \frac{d}{dr} R_1 \bigg|_0 = rR_2 \frac{d}{dr}(rR_1) - rR_1 R_2 \bigg|_0 \qquad (4.13)$$

Therefore Eq.(4.11) takes form

$$rR_1 \frac{d}{dr}(rR_2) - rR_2 \frac{d}{dr}(rR_1) \bigg|_0 + 2m(E_1 - E_2) \int_0^\infty R_1 R_2 r^2 dr = 0 \qquad (4.14)$$

from which we obtain finally equation for orthogonality condition

$$m(E_1 - E_2) \int_0^\infty R_2 R_1 r^2 dr = \frac{1}{2} \lim_{r \to 0} \left[ u_2(r) \frac{du_1(r)}{dr} - u_1 \frac{du_2(r)}{dr} \right] \qquad (4.15)$$

We see that (4.2) and (4.15) have the same right-hand sides. Or self-adjointness and orthogonality conditions are equal to the same expressions. Therefore because the self-adjoint operator has orthogonal eigenfunctions, requirement of orthogonality automatically provides self-adjointness of $H_R$, which means that this way provides realization of SAE procedure. It is an essence of so-called "pragmatic approach" [31], which is much simpler and gets the same results as the strong mathematical full SAE procedure, provided the fundamental condition (2.7) is not violated. Moreover this method is physically more transparent. Just this method had been used by Case in his well-known paper [22]. Notice that all above considerations are true only for the radial Hamiltonian operator $\hat{H}_R$, because for other operators proportionality like (4.2) and (4.15) does not arise.

## 5. SAE procedure for radial Hamiltonian in different cases

Let us now study in which cases are the right-hand sides of (4.3) and (4.15) vanishing. We must distinguish regular and transitive potentials.

In case of regular potentials (3.1), as was mentioned above, we retain only first, regular (or standard) solution at the origin $(C_2 = 0)$,

$$R_{st} \underset{r \to 0}{\sim} a_s r^{l+1} \qquad (5.1)$$

Calculating the r.-h.-sides of (4.3) and (4.15) by this function, we get zero. Therefore for regular potentials the radial Hamiltonian $\hat{H}_R$ is self-adjoint on regular solutions and it does not need SAE.

Contrary to this case, for transitive attractive (3.4) potential one has to retain the additional solution $R_{add} \underset{r \to 0}{\sim} a_a r^{-1/2 - F}$ as well, because there are no reasons to neglect it. Now for both solutions, the r.-h.-sides of (4.3) and (4.15) are not zero in general. Indeed they equal to

$$\text{R.-H.-Side of (4.15)} = P\left(a_1^{st} a_2^{add} - a_2^{st} a_1^{add}\right) \tag{5.2}$$

**Remark.** The case $P = 0$ must be considered separately, when the general solution of (3.8) behaves as

$$\lim_{r \to 0} R = a_{st} r^{-\frac{1}{2}} + a_{add} r^{-\frac{1}{2}} \ln r = u_{st} + u_{add} \tag{5.3}$$

Thus, instead of (4.15) one obtains

$$m(E_1 - E_2)\int_0^\infty R_2 R_1 r^2 dr = -\frac{1}{2}\left(a_1^{st} a_2^{add} - a_2^{st} a_1^{add}\right) \tag{5.4}$$

So retaining additional solution causes the breakdown of orthogonality condition and consequently $\hat{H}_R$ is no more a self-adjoint operator.

It is natural to ask – how to fulfil the orthogonality condition? It is clear, that in both $P \neq 0$ and $P = 0$ cases one must require

$$a_1^{st} a_2^{add} - a_2^{st} a_1^{add} = 0 \tag{5.5}$$

or equivalently

$$\frac{a_{1\,add}}{a_{1\,st}} = \frac{a_{2\,add}}{a_{2\,st}} \tag{5.6}$$

In this case the radial Hamiltonian $\hat{H}_R$ becomes a self-adjoint operator. This generalize the Case result [22], who considered only standard solution.

So it is necessary to introduce so called SAE parameter, which in our case may be defined as

$$\tau \equiv \frac{a_{add}}{a_{st}} \tag{5.7}$$

$\tau$ parameter is the same for all levels (for fixed orbital $l$ momentum) and is real for bound states.

From expressions (3.8) and (5.7) it is clear that we have three particular cases:

i). $a_{add} = 0$ $(\tau = 0)$. We keep only standard solutions.

ii). $a_{st} = 0$ $(\tau = \pm\infty)$. We keep only additional solutions.

iii). $\tau \neq 0, \pm\infty$. Solutions are neither "pure" standard nor "pure" additional.

In the last case this parameter becomes arbitrary one and it may be restricted only from some physical requirements. In other words the mathematical sets of quantum mechanics may not be enough without invoking of specific physical ideas.

*Comment:*

It must be noted that in some cases the fundamental condition (2.7) is ignored (see, e.g,[28-29]). In this cases, or if $u(0) \neq 0$, the left-hand side of (4.15) can be written as

$$m(E_1 - E_2)\int_0^\infty R_2 R_1 r^2 dr = \frac{1}{2}\lim_{r\to 0}\left[u_2(r)\frac{du_1(r)}{dr} - u_1(r)\frac{du_2(r)}{dr}\right] =$$
$$u_1(r)|_{r=0} u_2(r)|_{r=0}\left[\frac{1}{u_1(r)|_{r=0}}\frac{du_1(r)}{dr}\bigg|_{r=0} - \frac{1}{u_2(r)|_{r=0}}\frac{du_2(r)}{dr}\bigg|_{r=0}\right]$$
(5.8)

after that the SAE parameter is introduced for vanishing of r.-h.-s. of this equation in the following manner [28-29]

$$K = \frac{1}{u(r)_1|_{r=0}}\frac{d[u_1(r)]}{dr}\bigg|_{r=0}$$
(5.9)

while this consideration contradicts to condition (2.7), it is permissible from mathematical point of view, but is beyond the real physics as was explained in the introduction. As regards of the one dimensional Schrodinger equation, such a consideration is clearly permissible and valid [32]. While it must be underlined, that the consideration on the half-plane $(x \geq 0)$ doesn't correspond to the radial Schrodinger equation in 3-dimensions with $l = 0$.

### 6. "Fall" of a particle to the center

As a first application of retaining of additional solution let us reconsider the classical problem of particle's "falling to the center". It is described in many textbooks and is used in many articles. Most frequently, the book [23] is referenced. In this book, potential of kind (3.4) is regularized near the origin: in the range, $0 \leq r \leq r_0$ this potential is taken as constant and at the end, this regularization is removed $(r_0 \to 0)$. Using this procedure it is argued that the additional solution must be neglected in (3.8). However, because $R_{add} = a_{add} r^{-\frac{1}{2}-P}$ satisfies fundamental requirement (2.7) in the interval (3.9), as we think, this regularization and subsequent neglecting is not necessary. We can see it in an alternative way.

First let us make some remarks concerning to nodes of wave function. According to well-known theorem for the regular potentials (3.1) about the number of nodes for bound states (see, e.g. [20]), the n-th eigenfunction has n-1 nodes (or the ground state eigenfunction does not have nodes). It is easy to show that this theorem remains valid for the attractive potentials like (3.4). Besides that, the second theorem, according to which the number of bound states coincides with the number of nodes of Schrodinger wave function $R(r)$ in $E = 0$ state [20], is also valid for the potential (3.4). Below we consider examples, where these properties are applied.

Let us rewrite equation (3.7) (in resemblance to [23])

$$R'' + \frac{2}{r}R' + \frac{\gamma^2}{r^2}R = 0$$
(6.1)

where the constant

$$\gamma = 2mV_0 - l(l+1) \tag{6.2}$$

is related to above introduced $P$ as follows

$$P = \sqrt{1/4 - \gamma} \tag{6.3}$$

Let's search the solution of equation (6.1) in the form $R \sim r^s$. Then we find quadratic equation for s

$$s^2 + s + \gamma = 0 \tag{6.4}$$

with solutions

$$s_1 = -\frac{1}{2} + \sqrt{1/4 - \gamma}; \quad s_2 = -\frac{1}{2} - \sqrt{1/4 - \gamma} \tag{6.5}$$

Consider first the case $0 < \gamma < 1/4$ or $0 < P < 1/2$, when $s_1$ and $s_2$ are real numbers. Thus, the general solution of equation (6.1) should be

$$R = Ar^{s_1} + Br^{s_2} = Ar^{-\frac{1}{2}+P} + Br^{-\frac{1}{2}-P} = u_{st} + u_{add} \tag{6.6}$$

Here $u_{add}$ is more singular at the origin, than $u_{st}$, but in the interval (3.9) they both have the same properties and must be retained. As one saw in the previous section, this causes introduction of SAE procedure for Hamiltonian.

If $\gamma < 0$ or $P > 1/2$, one must keep only $R_{st}$.

When $\gamma > 1/4$, or P becomes imaginary number, then $s_1$ and $s_2$ should be mutually conjugated complex numbers

$$s_1 = -\frac{1}{2} + i\sqrt{\gamma - 1/4}; \quad s_2 = s_1^* \tag{6.7}$$

In this case the general solution of Eq. (6.1) will be

$$R \approx Ar^{-\frac{1}{2}+i\sqrt{\gamma-1/4}} + Br^{-\frac{1}{2}-i\sqrt{\gamma-1/4}} = Ar^{-\frac{1}{2}} \exp\left[i\left(\sqrt{\gamma-1/4}\ln r\right)\right] + Br^{-\frac{1}{2}} \exp\left[-i\left(\sqrt{\gamma-1/4}\ln r\right)\right] \tag{6.8}$$

We see that both solutions oscillate and have same singularity at origin. Taking into account that for for bound states the wave function R must be real, we are forced to require $B^* = A$ and therefore

$$R \approx Ar^{-\frac{1}{2}} \cos\left(\sqrt{\gamma - 1/4}\ln r + \alpha\right) \tag{6.9}$$

where $\alpha$ is an arbitrary constant – the phase of $B$ relative to $A$. Therefore retaining of both solutions causes introduction of "superfluous" parameter $\alpha$, which really is a SAE parameter [33]. If we follow the discussion given in [23], we can show that wave function (6.9) corresponds to "falling to the center".

Therefore, it is evident that if we retain $R_{add}$ in $0 < \gamma < 1/4$ domain ($0 < P < 1/2$), the problem of "falling to the center" can be considered without modification (regularization) of potential. It is just the alternate view to this problem.

Morever, one can easily confirm that in case $\gamma > 1/4$, the requirement of finiteness of kinetic energy gives the following limitation $\text{Re}\, s_{1,2} > -\frac{1}{2}$, but now $\text{Re}\, s_{1,2} = -\frac{1}{2}$. Therefore, in this case both solutions have the same behavior and give infinite kinetic energy. Thus, the argument of authors in Ref. [24,25] against the additional solution fails.

## 7. What is new for Inverse square potential when we retain additional solutions?

Consider the following potential

$$V = -\frac{V_0}{r^2}, \qquad V_0 > 0 \tag{7.1}$$

in the whole space. There is only one worthy case, namely $0 < P < 1/2$.

Now the wave function R for $E = 0$ has the form (6.6) in the whole space. It has a single zero, determined by

$$r_0 = \left(-\frac{B}{A}\right)^{1/2P} \tag{7.2}$$

(It is evident from this relation that constants A and B must have opposite signs in order for $r_0$ to be real number). Therefore, the wave function has only one node and according to well-known theorem we have one bound state only. This result differs from that considered in any textbooks of quantum mechanics.

We can give very simple physical picture of how the additional solutions arise. For this purpose, let us rewrite the Schrodinger equation near the origin for attractive potential (3.4) in the form

$$R'' + \frac{2}{r}R' + 2m[E - V_{ac}(r)]R = 0 \tag{7.3}$$

where

$$V_{ac} = \frac{P^2 - 1/4}{2mr^2} \tag{7.4}$$

Consider the following possible cases:

i). If $P > 1/2$, then $V_{ac} > 0$ and it is repulsive centrifugal potential and as we saw, one has no additional solutions.

ii). If $0 < P < 1/2$, then $V_{ac} < 0$. Therefore, it becomes attractive and is called as quantum anti-centrifugal potential [34]. This potential has $R_{add}$ states, because the condition (2.7) is fulfilled in this case.

iii). If $P^2 < 0,$ then $V_{ac}$ becomes strongly attractive and one has "falling to the center".

Therefore, the sign of the potential $V_{ac}$ determines whether we need additional solutions or not.

It was thought that potential (7.1) had no levels out of region of "falling to the center" (See e.g. [22,23]), but in [5,35,36] single level was found by complete SAE procedure, while the boundary condition and the range of parameter, like P are questionable there. Here we'll show explicitly that this potential has exactly a single level, which depends on the SAE parameter $\tau$.

Let's take the Schrodinger equation for potential (7.1)

$$\frac{d^2 R}{dr^2} + \frac{2}{r}\frac{dR}{dr} + \left(-k^2 - \frac{P^2 - 1/4}{r^2}\right)R = 0 \tag{7.5}$$

where P is given by (3.7) and

$$k^2 = -2mE > 0; \quad (E < 0) \tag{7.6}$$

One can reduce Eq.(7.5) to the equation for modified Bessel functions by substitutions

$$R(r) = \frac{f(r)}{\sqrt{r}}; \quad x = kr \tag{7.7}$$

leading to the following equation

$$x^2 \frac{d^2 f(x)}{dx^2} + x\frac{df(x)}{dx} - \left(x^2 + P^2\right)f(x) = 0 \tag{7.8}$$

This equation has 3 pairs of independent solutions: $I_P(kr)$ and $I_{-P}(kr)$, $I_P(kr)$ and $e^{i\pi P}K_P(kr)$, $I_{-P}(kr)$ and $e^{i\pi P}K_P(kr)$, where $I_P(kr)$ and $K_P(kr)$ are Bessel and MacDonald modified functions [37], respectively. Consider these possibilities separately.

1) The pair $I_P(kr)$ and $I_{-P}(kr)$:

The general solution of (7.5) is

$$R = r^{-\frac{1}{2}}\left[AI_P(kr) + BI_{-P}(kr)\right] \tag{7.9}$$

Consider the behaviour of this solution at small and large distances:

a) Small distances

In this case [37]

$$I_P(z) \underset{z\to 0}{\approx} \left(\frac{z}{2}\right)^P \frac{1}{\Gamma(P+1)} \tag{7.10}$$

Then it follows from (7.9) and (7.10) that

$$\lim_{r\to 0} R(r) \approx r^{-\frac{1}{2}}\left[A\left(\frac{k}{2}\right)^P \frac{r^P}{\Gamma(P+1)} + B\left(\frac{k}{2}\right)^{-P} \frac{r^{-P}}{\Gamma(1-P)}\right] \tag{7.11}$$

From (3.8), (7.11) and the definition (5.7) we obtain $\tau$,

$$\tau = \frac{B}{A} 2^{2P} k^{-2P} \frac{\Gamma(1+P)}{\Gamma(1-P)} \qquad (7.12)$$

b) Large distances

In this case [37]

$$I_P(z) \underset{z \to \infty}{\approx} \frac{e^z}{\sqrt{2\pi z}} \qquad (7.13)$$

and

$$R(r) \underset{r \to \infty}{\approx} \frac{1}{\sqrt{2\pi}} \{A + B\} e^{kr} \qquad (7.14)$$

Therefore, requiring vanishing of $R(r)$ at infinity, we have to take

$$B = -A \qquad (7.15)$$

and from (7.12), (7.15) and (7.6) we obtain one real level (for fixed orbital $l$ momentum, satisfying (3.5)),

$$E = -\frac{2}{m} \left[ \frac{\Gamma(1+P)}{\Gamma(1-P)} \right]^{\frac{1}{P}} \left[ -\frac{1}{\tau} \right]^{\frac{1}{P}} ; \quad 0 < P < 1/2 \qquad (7.16)$$

Eq. (7.16) is a new expression derived as a consequence of orthogonality condition in the framework of "pragmatic" approach. It differs from the form obtained in [5,35,36], where the complete SAE was used, while the boundary codition and the range of parameter, like P, are questionable.

Reality of energy in (7.16) restricts $\tau$ parameter to be negative $\tau < 0$. In general $\tau$ is a free parameter but some physical requirements may restrict its magnitude. Note that this level is absent in standard quantum mechanics without SAE procedure.

To obtain corresponding wave function, take into account a well-known relation [37]

$$K_P(z) = \frac{\pi}{2 \sin P\pi} [I_{-P}(z) - I_P(z)] \qquad (7.17)$$

Then the wave function corresponding to the level (7.16) is

$$R = -A \frac{2}{\pi} r^{-\frac{1}{2}} \sin P\pi \cdot K_P(kr) \qquad (7.18)$$

Because of exponential damping

$$K_P(z) \underset{z \to \infty}{\approx} \sqrt{\frac{\pi}{2z}} e^{-z} \qquad (7.19)$$

the function (7.18) corresponds to the bound state. It is also known that $K_P(z)$ function has no zeroes for real P $(0 < P < 1/2)$ and therefore (6.14) corresponds to single bound state. Moreover, wave function (7.18) satisfies the fundamental condition (2.7) for $0 < P < 1/2$.

2) The pair $I_P(kr)$ and $e^{i\pi P} K_P(kr)$;

The general solution of (7.5) is

$$R = r^{-\frac{1}{2}} \left[ A I_P(kr) + B e^{i\pi P} K_P(kr) \right] \tag{7.20}$$

At large distances

$$\lim_{r \to \infty} R(r) \approx \frac{1}{\sqrt{2\pi}} \left( A e^{kr} + B e^{i\pi P} \pi e^{-kr} \right) \approx A \frac{e^{kr}}{\sqrt{2\pi}} \tag{7.21}$$

Therefore we have no bound states.

The same follows for pair $I_{-P}(kr)$ and $e^{i\pi P} K_P(kr)$. Thus only pair $I_P(kr)$ and $I_{-P}(kr)$ has a single bound state.

Noting that the considatation of all possible pairs of solution is, in general, necessary, because there is no guide principle, by which one can guess which pair must be considered.

Let us make some comments, concerning to the application of above results.

a) Owing to the fact that the Schrodinger equation has a single level for inverse square potential after SAE procedure, one can make some comments about monopole problem where exactly like potential (7.1) is applied. Contrary to common opinion there may appear new bound state solutions after a self-adjoint extention. This point will be discussed elsewhere.

b) In [35] it was noticed that single bound state may be observed experimentally in polar molecules. For example, $H_2S$ and HCl exhibit anomalous electron scattering [38,39], which can be explained only by electron capture. Indeed, for those molecules electron is moving in a point dipole field, and, in this case the problem is reduced to the Schrodinger equation with a potential (7.1). Thus, a level (7.16) obtained theoretically may be observed in those experiments.

c) It was commonly believed, that the potential

$$V = -\frac{V_0}{sh^2 \alpha r} \tag{7.22}$$

has no levels in (3.10) region (see for example problem 4.39 in [40]). In [40] by the arguments of well-known comparison theorem [30], which in this case looks like

$$-\frac{V_0}{sh^2 \alpha r} \geq -\frac{V_0}{\alpha^2 r^2} \tag{7.23}$$

it is concluded that the potential (7.22) can not have a level in the area (3.10), because the potential (7.1) has no levels in this area. But, as we know, there is (7.16) $\tau$ depended one level, therefore the levels for (7.22) are expected. Indeed, in [41] by using the Nikiforov-Uvarov method [42], it is shown that (7.22) potential has infinite number of levels in the (3.10) region.

Now let us turn to more realistic models.

## 8. The valence electron model

It is well known that the potential

$$V = -\frac{V_0}{r^2} - \frac{\alpha}{r}; \quad (V_0, \alpha > 0) \tag{8.1}$$

is used for the description of alkaline metal (Li,Na,K,Rb,Cs) atoms' spectra [5]. Add to this the similar potential "naturally" arises in the Klein – Gordon equation for the Coulomb interaction, for which SAE will be discussed below.

This potential, unlike to the Coulomb one, has a singular $r^{-2}$ like behavior at the origin. Therefore according to our strategy one must consider equation for the $R(r)$ function, which in dimensionless variables takes form

$$\left(\frac{d^2}{d\rho^2} + \frac{2}{\rho}\frac{d}{d\rho} - \frac{P^2 - 1/4}{\rho^2} + \frac{\lambda}{\rho} - \frac{1}{4}\right)R = 0 \tag{8.2}$$

where

$$\rho = \sqrt{-8mE}\,r = ar; \quad \lambda = \frac{2m\alpha}{\sqrt{-8mE}} > 0, \quad E < 0 \tag{8.3}$$

and $P$ is again done by Eq.(3.7).

If we use the notation of [43],

$$R = \rho^{-\frac{1}{2}+P} e^{-\frac{\rho}{2}} F(\rho), \tag{8.4}$$

the equation for confluent hypergeometric functions follows

$$\rho F'' + (2P + 1 - \rho)F' - (1/2 + P - \lambda)F = 0 \tag{8.5}$$

This equation has four independent solutions, two of which constitute a fundamental system of solutions [44]. They are (in notations of [44]):

$$\begin{aligned} y_1 &= F(a,b;\rho) \\ y_2 &= \rho^{1-b} F(1+a-b, 2-b; \rho) \\ y_5 &= \Psi(a,b;\rho) \\ y_7 &= e^{\rho} \Psi(b-a, b; -\rho) \end{aligned} \tag{8.6}$$

where

$$a = 1/2 + P - \lambda, \quad b = 1 + 2P \tag{8.7}$$

Only $y_1$ is considered in the scientific articles, as well as in all textbooks (see, e.g. [5, 23]). Requiring $a = -n$ ($n = 0,1,2,...$) the standard levels follow. Other solutions $(y_2, y_5, y_7)$ have singular behavior at the origin and usually they are not taken into account. But as was mentioned frequentative, the singularity in case of attractive potentials like (3.4) has the form $r^{-\frac{1}{2}-P}$ and in the region $0 < P < 1/2$ other solutions must be considered as well. Therefore, the problem becomes more "rich".

Let us consider a pair $y_1$ and $y_2$. The general solution of (8.5) is

$$R = C_1 \rho^{-1/2+P} e^{-\frac{\rho}{2}} F(1/2 + P - \lambda, 1 + 2P; \rho)$$
$$+ C_2 \rho^{-1/2-P} e^{-\frac{\rho}{2}} F(1/2 - P - \lambda, 1 - 2P; \rho) \qquad (8.8)$$

From the behavior of (8.8) at the origin and from (5.7), we obtain the following expression for SAE $\tau$ parameter

$$\tau = \frac{C_2}{C_1} \frac{1}{(-8mE)^P} \qquad (8.9)$$

Note on the other hand that, R must decrease at infinity. From well-known asymptotic properties of confluent hypergeometric function F, we find the following restriction

$$C_1 \frac{\Gamma(1+2P)}{\Gamma(1/2+P-\lambda)} + C_2 \frac{\Gamma(1-2P)}{\Gamma(1/2-P-\lambda)} = 0 \qquad (8.10)$$

It gives an equation for eigenvalues in terms of $\tau$ parameter

$$\frac{\Gamma(1/2 - \lambda - P)}{\Gamma(1/2 - \lambda + P)} = -\tau(-8mE)^P \frac{\Gamma(1-2P)}{\Gamma(1+2P)} \qquad (8.11)$$

We see that this is very complicated transcendental equation for E, depending on $\tau$ parameter. There are two values of $\tau$, when this equation can be solved analytically:

i) $\tau = 0$. In this case we have only standard levels, which can be found from the condition that $\Gamma(1/2 - \lambda + P)$ has poles

$$1/2 - \lambda + P = -n_r; \quad n_r = 0,1,2... \qquad (8.12)$$

ii) $\tau = \pm\infty$. In this case we have only additional levels, obtained from the poles of $\Gamma(1/2 - \lambda - P)$

$$1/2 - \lambda - P = -n_r; \quad n_r = 0,1,2... \qquad (8.13)$$

Thus, in these cases i) and ii) one can obtain explicit expression for standard and additional levels

$$E_{st,add} = -\frac{m\alpha^2}{2[1/2 + n_r \pm P]^2} = -\frac{m\alpha^2}{2\left[1/2 + n_r \pm \sqrt{(l+1/2)^2 - 2mV_0}\right]^2} \qquad (8.14)$$

where signs (+) or (–) correspond to standard and additional levels, respectively.

iii) For arbitrary $\tau$ parameter the equation (8.11) is discussed in the Appendix A of [15].

Let us study the asymptotics of equation (8.11) for large values of $\lambda$, which allows us approximately find explicit dependence of E on $\tau$. As it is evident from (8.11), arguments of $\Gamma$ functions are negative for large values of $\lambda$. Therefore if we reflect the signs of arguments with the aid of well-known relation [37]

$$\Gamma(z)\Gamma(-z) = -\frac{1}{z}\frac{\pi}{\sin(\pi z)} \tag{8.15}$$

we find the approximate *dependence of energy on $\tau$ parameter*

$$E \approx -\frac{m\alpha^2}{8P^2}\left[1+\frac{\Gamma(1-2P)}{\Gamma(1+2P)}(2m\alpha)^{2P}\tau\right]^2 \tag{8.16}$$

We note that only the Eq. (8.12) was known till now. So the equation (8.11) and its consequences are new results.

Notice also that, if we take $V_0 < 0$ in (8.1), then we obtain well-known Kratzer potential [5], but in this case the condition (3.5) is not satisfied. Therefore there are no additional levels for Kratzer potential.

In monograph [5] energy levels for alkaline metal atoms are written in Ballmer's form

$$E_{n'} = -R\frac{1}{n'^2} \tag{8.17}$$

where R is a Rydberg constant and $n'$ is the effective principal quantum number

$$n' = n_r + l' + 1 \qquad (n_r = 0,1,2...) \tag{8.18}$$

$l'$ is defined from equation

$$l'(l'+1) = l(l+1) - 8mV_0 \tag{8.19}$$

or

$$l' = -1/2 \pm P = -1/2 \pm \sqrt{(l+1/2)^2 - 2mV_0} \tag{8.20}$$

Only (+) sign was considered in front of the square root until now. In [5] $V_0$ was considered to be small and after expansion of this root, approximate expression for the standard levels was derived

$$E_{st} = -R\frac{1}{(n+\Delta_l)^2}; \quad n = n_r + l + 1 \tag{8.21}$$

where

$$\Delta_l \equiv \Delta_l^{st} = -\frac{2mV_0}{2l+1} \tag{8.22}$$

is so - called Rydberg correction (quantum defect) [5].

As regards of additional levels, this procedure is invalid, because $V_0$ is bounded from below according to (3.5).

Aapproximate expansion for additional levels is possible only for $l = 0$. We have in this case

$$P = \sqrt{\frac{1}{4} - 2mV_0} \approx \frac{1}{2}(1 - 4mV_0) \tag{8.23}$$

$V_0$ may be arbitrary small, but different from zero, because in this case $P = 1/2$ and we have no additional levels.

Let us rewrite now the function (8.8) in united form by using the following relation for the Whittaker functions [44]

$$W_{a,b}(x) = e^{-\frac{1}{2}x} x^{\frac{1}{2}+b} \frac{\pi}{\sin \pi(1+2b)} \left[ \frac{F(1/2+b-a, 1+2b; x)}{\Gamma(1/2-a-b)\Gamma(1+2b)} - x^{-2P} \frac{F(1/2-a-b, 1-2b; x)}{\Gamma(1/2+b-a)\Gamma(1-2b)} \right] \quad (8.24)$$

Then from (8.3), (8.8), (8.10) and (8.24) we derive

$$R(r) = C_1 \Gamma(1+2P)\Gamma(1/2-P-\lambda) \frac{\sin \pi(1+2P)}{\pi r} W_{\lambda,P}\left(\sqrt{-8mE}r\right) \quad (8.25)$$

Because the Whittaker function $W_{a,b}(x)$ has an exponential damping [45]

$$W_{a,b}(x) \underset{x \to \infty}{\approx} e^{-\frac{1}{2}x} x^a, \quad (8.26)$$

it is clear that (8.25) corresponds to a bound state. Moreover, it satisfies the fundamental condition (2.7) for $0 < P < 1/2$ interval.

Therefore, for $\tau = 0, \pm\infty$ the standard and additional levels are obtained from (8.14) with corresponding wave functions

$$R_{st} = C_1 \rho^{-1/2+P} e^{-\frac{\rho}{2}} F(1/2+P-\lambda, 1+2P; \rho) \quad (8.27)$$

$$R_{add} = C_2 \rho^{-1/2-P} e^{-\frac{\rho}{2}} F(1/2-P-\lambda, 1-2P; \rho) \quad (8.28)$$

For arbitrary $\tau \neq 0, \pm\infty$ the energy can be obtained from the transcendental equation (8.11), while the wave function is given by (8.25).

The united form (8.25) is also new and it is a consequence of the SAE procedure.

According to [44] our function (8.25) takes the following form

$$R(r) = C_1 \Gamma(1+2P)\Gamma(1/2-P-\lambda) \frac{\sin \pi(1+2P)}{\pi \rho} e^{-\frac{\rho}{2}} \rho^{\frac{1}{2}-P} \Psi\left(\frac{1}{2}-\lambda-P, 1-2P; \rho\right) \quad (8.29)$$

where $\Psi(a,b,x)$ is one of the above mentioned solutions, (8.6), namely $y_5$. Its zeros are well-studied [44]: For real $a, b$ (note that in our case $a = \frac{1}{2} - \lambda - P$; $b = 1 - 2P$ are real numbers) this function has finite numbers of positive roots. However, for the ground state there are three cases where this function has no zeros:

1) $a > 0$; 2) $a - b + 1 > 0$; 3) $-1 < a < 0$ and $0 < b < 1$. Only the last case is interesting for us, because $a = \frac{1}{2} - \lambda - P$; $b = 1 - 2P$ and P is in the interval (3.9). It means

$$-1 < 1/2 - P - \frac{2m\alpha}{\sqrt{-8mE}} < 0 \qquad (8.30)$$

In other words, the ground state energy, which is given by transcendental equation (8.11), must obey this inequality.

The wave function in form of (8.29) is new.

Let us now make some comments:

I) One can easily obtain the existency condition of additional levels from (8.21) and (3.5) in diverse form

$$l < \Delta_l < l + 1 \qquad (8.31)$$

If we use data of monograph [5], we obtain that for $l = 0$ states only Li, for $l = 1$ only Ka and for $l = 2$ only Cs satisfy (8.31) (i.e. they have additional solutions and it is necessary to carry out SAE procedure), and Na and Rb have no additional levels. The condition (8.31) is also new, which helps us to determine which alkaline metals need SAE extension of Hamiltonian.

II) We have following situation in case of choosing another pairs of solutions of (8.6):

1) ($y_5$ and $y_7$) - do not have levels.

2) ($y_1$ and $y_5$) - give only standard levels (nothing new).

3) ($y_2$ and $y_5$) - give only pure additional levels ($\tau = \pm\infty$), which is unjustified physically, because the standard levels are completely lost.

4) ($y_2$ and $y_7$) - not permissible, because in this case $\tau = 0$ is forbidden and we have no standard levels.

5) ($y_1$ and $y_7$) - not allowed, because in limit $\alpha \to 0$ no levels follow for potential $V = -\frac{V_0}{r^2}$, but as we've seen above there exists a single level for this potential.

**9. Klein-Gordon equation with Coulomb potential and "hydrino"**

We note that the problems of additional levels were discussed by other authors as well [45-48]. In particular, in [45] the Klein – Gordon equation is considered with $V = -\frac{\alpha}{r}$ Coulomb potential

$$R'' + \frac{2}{r}R' + \left[E^2 - m^2 - \frac{l(l+1)}{r^2} + \frac{2E\alpha}{r} + \frac{\alpha^2}{r^2}\right]R = 0 \qquad (9.1)$$

The author underlines, that there must be levels below the standard levels (called, "hydrino" eigenstates).

Let consider this problem in more detail. First of all note that the equation (9.1) coincides with Eq. (8.2), but now

$$\rho = 2\sqrt{m^2 - E^2}; \quad \lambda = \frac{E\alpha}{\sqrt{m^2 - E^2}}; \quad P = \sqrt{(l+1/2)^2 - \alpha^2} > 0 \tag{9.2}$$

We must require $m^2 > E^2$ for bound states. Therefore one can use all the previous relations from valence electron model taking into account the definitions (9.2). In particular the SAE parameter now is

$$\tau = \frac{C_2}{C_1} \frac{1}{\left(2\sqrt{m^2 - E^2}\right)^P} \tag{9.3}$$

and for eigenstates we have the following equation

$$\frac{\Gamma(1/2 - \lambda - P)}{\Gamma(1/2 - \lambda + P)} = -\tau \left(2\sqrt{m^2 - E^2}\right)^P \frac{\Gamma(1 - 2P)}{\Gamma(1 + 2P)} \tag{9.4}$$

from which for $\tau = 0$ and $\tau = \pm\infty$ we derive the standard and additional levels in analogy of Eq. (8.14)

$$E_{st} = \frac{m}{\sqrt{1 + \frac{\alpha^2}{(1/2 + n_r + P)^2}}}; \quad n_r = 0,1,2... \tag{9.5}$$

$$E_{add} = \frac{m}{\sqrt{1 + \frac{\alpha^2}{(1/2 + n_r - P)^2}}}; \quad n_r = 0,1,2... \tag{9.6}$$

Exactly these (9.6) levels are called a "hydrino" levels in [45-48]. It is evident that the hydrino levels are analogical to $E_{add}$ states Eq.( 8.14), but these two cases differ from each others. Particularly, it is possible to pass the limit $V_0 \to 0$ in the equation (8.2) and obtain Hydrogen's problem. Usually this limiting procedure is used in traditional textbooks to choose between two signs in (8.12), while in (9.1) constants for both terms in potential are mutually proportional ($\alpha$ and $\alpha^2$), and vanishing of one of them causes vanishing of another, so we turn to the free particle problem insread of Coulomb one. Moreover in those papers [45-48] the SAE procedure was not used. They considered only two signs in front of square root in equation analogous to (8.18) and only (9.5) and (9.6) levels are considered, which correspond only to cases $\tau = 0$ and $\tau = \pm\infty$. Contrary to that case we performed SAE procedure and take attention to the hydrino problem, in case of $\tau = \pm\infty$.

The difference between standard and hydrino states manifests clearly in the nonrelativistic limit when $\alpha \to 0$. Indeed, let write the relations (9.5-6) for the ground state ($n_r = l = 0$):

$$E_{st}^{(0)} = \frac{m}{\sqrt{2}} \sqrt{1 + \sqrt{1 - 4\alpha^2}} \tag{9.7}$$

$$E_{Hyd} \equiv E_{add}^{(0)} = \frac{m}{\sqrt{2}}\sqrt{1-\sqrt{1-4\alpha^2}} \tag{9.8}$$

Expantion in powers of $\alpha$ gives

$$E_{st}^{(0)} = m\left(1-\frac{\alpha^2}{2}\right) \tag{9.9}$$

$$E_{HYD}^{(0)} = m\alpha \tag{9.10}$$

Let us note that if we expand $l = 0; n_r \neq 0$ states till to order of $\alpha^2$ it follows

$$E_{st}^{(0)} = m\left(1-\frac{\alpha^2}{2(n_r+1)^2}\right) \tag{9.11}$$

$$E_{HYD}^{(0)} = m\left(1-\frac{\alpha^2}{2(n_r)^2}\right) \tag{9.12}$$

Comparision of which shows that we have some kind of degeneracies between the levels with $n_r+1$ nodes of hydrino and energies for $n_r$ nodes of standard states.

In [48] it is noted that the hydrino states can be excluding by orthogonality requirement, but it is not correct. Detailed study considered in Appendix B of [15] shows that the hydrino states must be retained,

H.W.Crater et all. [3] considered relativistic quasipotential approach to the magnetic resonance problem at short distances. Retaining the additional singular solution they derived hydrino states, which they called as "peculiar" states. They pick up the following important questions: Can the peculiar states be observed? Will there be transitions between the usual and peculiar states? … The fact that peculiar states of $(n+1)$ th $^1S_0$ state is nearly degenerate with the usual $n$ th $^1S_0$ state may facilitate such tunneling transition.

## 10. Conclusions

In this paper we have studied the inverse square potential in the framework of equation for the full radial function $R(r)$ exploring its boundary behavior near the origin, established by us in [1,2]. We have shown that there are no reasons to neglect the singular solution and therefore we retain it. We have investigated the possibility of realization of SAE procedure in the pragmatic approach, basing on orthogonality property of solutions under consideration. This procedure introduces an extra parameter $\tau$.

- We emphasize that after performing of SAE procedure $r^{-2}$ like behavioring potentials get one level of bound state. We derived this both on general framework and by explicit solution of corresponding equation. It is natural, that the energy eigenvalue depends on the SAE parameter $\tau$.

- In parallel of this we discussed analogous problems, considered by other authors. Our result differs from their mainly in that that we retained a non-regular solution. Moreover some differences result from the difference of the areas of parameters as a consequence of used boundary conditions.

- Consideration of particle's falling to the center is a peculiar example, when regularization of $r^{-2}$ term in potential is avoided in this case by inclusion of both solutions and performing SAE procedure. The regularization in this example is a particular case of SAE

- It must be underlined that obtained results are depending on the SAE parameter $\tau$ which is arbitrary. It is natural that in particular cases this parameter could be determined in accordance of considered physical problems.

- We considered also physically quite realistic examples, such as a valence electron model and relativistic Klein-Gordon equation with the Coulomb potential, which is considerably related to the previous example.

- We obtain a solution for the radial wave function in terms of special functions and present a united expression for this solution, from which the separate cases follow. The problem of zeros of this function is also investigated and the definite predictions about the levels of alkaline metal atoms are presented.

At last, we connect this general model to the Klen-Gordon equation with the Coulomb potential retaining also the non-regular solution. We have shown that this additional solution has no Balmer's like nonrelativistic limit. But appearing here new states called "hydrino" or "peculiar" states may be in principle observable by tunneling transition into the usual states. This phenomenon may happen only in $^1S_0$ (or $l = 0$ state, according to inequalities derived in our paper).

Notice at last that many related problems are considered by us in the earlier paper [15], which was based on the considered above boundary condition $u(0) = 0$, while that time we did not know, that this restriction is so strong and important. More relevant references may be found also in that paper.

**Acknowledgments**

The authors thank Drs. G.Japaridze T.Kereselidze, A.Kvinikhidze M.Nioradze and participants of seminars at Iv. Javakhishvili Tbilisi State University, for many valuable comments and discussions. Authors acknowledges financial support of the Shota Rustaveli National Science Foundation (Projects DI/13/02 and FR/11/24)